\documentclass[useAMS,usenatbib,usegraphicx]{mn2e}
\usepackage[pdftex,colorlinks,citecolor=blue,linkcolor=blue,pdfauthor= Julian~Merten,pdftitle =Abell~2744,pdfsubject=Cosmology]{hyperref}
\usepackage{amssymb}
\def\aj{AJ}%
\def\araa{ARA\&A}%
\def\apj{ApJ}%
\def\apjl{ApJ}%
\def\apjs{ApJS}%
\def\aap{A\&A}%
\def\mnras{MNRAS}%
\def\na{New A}%
\def\pasp{PASP}%
\def\pasj{PASJ}%
\def\gca{Geochim.~Cosmochim.~Acta}%
\def\physrep{Phys.~Rep.}%
\def\simgt{\lower.5ex\hbox{$\; \buildrel > \over \sim \;$}}

\begin{document}


\title[The galaxy cluster merger Abell~2744]{
Creation of cosmic structure in the complex galaxy cluster merger Abell~2744}
\author[J.~Merten~et~al.]
{
J.~Merten,$^{1,2}$\thanks{Both authors contributed equally to this work}\thanks{E-mail:~jmerten@ita.uni-heidelberg.de}
D.~Coe,$^{3}$\footnotemark[1]
R.~Dupke,$^{4,5,6}$ 
R.~Massey,$^{7}$
A.~Zitrin,$^{8}$
E.~S.~Cypriano,$^{9}$
\newauthor
N.~Okabe,$^{10}$
B.~Frye,$^{11}$
F.~G.~Braglia,$^{12}$
Y.~Jim\'enez-Teja,$^{13}$
N.~Ben\'{i}tez,$^{13}$
\newauthor
T.~Broadhurst,$^{14,15}$
J.~Rhodes,$^{16,17}$
M.~Meneghetti,$^{2}$
L.~A.~Moustakas,$^{17}$
L.~Sodr\'e Jr.,$^{9}$
\newauthor
J.~Krick$^{18}$
and J.~N.~Bregman$^{4}$\\
$^{1}$Institut f\"{u}r Theoretische Astrophysik, ZAH, Albert-Ueberle-Stra\ss e 2, 69120 Heidelberg, Germany\\
$^{2}$INAF-Osservatorio Astronomico di Bologna, Via Ranzani 1, 40127 Bologna, Italy \\
$^{3}$Space Telescope Science Institute, Baltimore, MD 21218, USA\\
$^{4}$University of Michigan, Ann Arbor, MI 48109-1090, USA \\
$^{5}$Eureka Scientific, 2452 Delmer Street 100, Oakland CA 94602-3017, USA\\
$^{6}$Observat\'{o}rio Nacional, Rua General Jos\'{e} Cristino 77, Rio de Janeiro, 20921-400, Brazil\\
$^{7}$University of Edinburgh, Royal Observatory, Blackford Hill, Edinburgh EH9 3HJ, UK \\
$^{8}$School of Physics and Astronomy, Raymond and Beverly Sackler Faculty of Exact Sciences, Tel Aviv University, Tel Aviv 69978, Israel \\
$^{9}$Instituto de Astronomia, Geof\'isica e Ci\^encias Atmosfericos, Universidade de S\~ao Paulo (IAG/USP), 05508-090 S\~ao Paulo/S.P.,Brazil\\
$^{10}$Academia Sinica Institute of Astronomy and Astrophysics (ASIAA), P.O. Box 23-141, Taipei 10617, Taiwan\\
(Additional affiliations after the conclusions)
}
\maketitle
\begin{abstract}
We present a detailed strong lensing, weak lensing and X-ray analysis of Abell~2744 ($z = 0.308$), 
one of the most actively merging galaxy clusters known.  
It appears to have unleashed `dark', `ghost', `bullet' and `stripped' substructures, 
each $\sim10^{14} M_{\odot}$.
The phenomenology is complex and will present a challenge for numerical simulations to reproduce.
With new, multiband {\sl HST} imaging, 
we identify 34 strongly-lensed images of 11 galaxies around the massive Southern `core'.
Combining this with weak lensing data from {\sl HST}, {\sl VLT} and {\sl Subaru}, 
we produce the most detailed mass map of this cluster to date.
We also perform an independent analysis of archival {\sl Chandra} X-ray imaging.
Our analyses support a recent claim that the Southern core and Northwestern substructure
are post-merger and exhibit morphology similar to the \textit{Bullet Cluster} viewed from an angle.
From the separation between X-ray emitting gas and lensing mass in the Southern core,
we derive a new and independent constraint on the self-interaction cross section of dark matter particles $\sigma/m < 3 \pm 1~\textrm{cm}^{2} / \textrm{g}$.
In the Northwestern substructure, the gas, dark matter, and galaxy components
have become separated by much larger distances.
Most curiously, the `ghost' clump (primarily gas) 
\textit{leads} the `dark' clump (primarily dark matter) by more than $150$~kpc.
We propose an enhanced `ram-pressure slingshot' scenario
which may have yielded this reversal of components with such a large separation, but needs further confirmation by follow-up observations
and numerical simulations.
A secondary merger involves a second `bullet' clump in the North 
and an extremely `stripped' clump to the West.
The latter appears to exhibit the largest separation between dark matter and X-ray emitting baryons detected to date in our sky.
\end{abstract}

\begin{keywords}
gravitational lensing: strong --- gravitational lensing: weak --- galaxies: clusters: individual: Abell~2744 --- dark matter --- large-scale structure of the Universe ---   X-rays: individual: Abell~2744.
\end{keywords}

\section{Introduction}
The standard $\Lambda$CDM cosmological model suggests a bottom-up sequence of structure formation,
in which a series of merging events culminates in massive clusters of galaxies, the latest structures to form in the observable Universe \citep{Bond1991,Lacey1993}. 
The number of clusters as a function of their mass (the steep, high end of the mass function) depends sensitively upon cosmological parameters \citep[e.g.][]{Vikhlinin2009} 
and has become an important observational test of cosmology.
Most measurements of cluster masses rely upon the calibration of more easily observable proxies, such as X-ray luminosity, temperature or galaxy richness.
However, clusters form through multiple, dynamic accretions, so are likely to be turbulent places, and the turmoil affects those observable proxies.
It is therefore vital to quantitatively understand the merging process, for example by mapping the distribution of dark matter, stars and baryonic gas in
systems at many different stages of the merger.

Merging clusters of galaxies have become useful laboratories in which to study the nature and interaction properties of dark matter.
The best-studied example is 
the \textit{Bullet Cluster} 1ES~0657-558 ($z=0.296$) \citep{Tucker1998,Markevitch2002}. Combined X-ray and gravitational lensing analyses
show a clear separation between the centres of X-ray emission and the peaks in surface mass density -- indicating a fundamental difference between
baryonic gas, which feels the pressure of the collision, and dark matter, which is nearly collisionless \citep{Clowe2004,Clowe2006,Bradav2006}. 
The discovery and interpretation of the \textit{Bullet Cluster} has inspired a lively debate about whether
such a system could exist in different cosmological models \citep[see e.g.][]{Hayashi2006}.
Improvements are continuing in numerical simulations \citep{Milosavljevi'c2007,Springel2007,Mastropietro2008,Lee2010}.
Observations have also broadened, with discoveries of a possible line-of-sight merger CL0024+1654 
\citep{Czoske2002,Hoekstra2007,Jee2007,Zitrin2009,Umetsu2010,Zu2009,Zu2009a},
and other systems including the \textit{Baby Bullet} (MACS~J0025.4-1222) \citep{Bradav2008}, 
the \textit{Cosmic Train Wreck} (Abell~520) \citep{Mahdavi2007,Okabe2008}, Abell~2146 \citep{Russell2010}, 
Abell~521 \citep{Giacintucci2008,Okabe2010a} and Abell~3667 \citep{Finoguenov2010}. 
All these systems place potentially tight constraints on the interaction between baryons and dark matter,
and are exemplary probes for our understanding of structure formation within gravitationally bound systems.

In the archival data of 38 merging clusters, \citet{Shan2010} found the
largest offset between X-ray and lensing signals to occur in the massive ($L_X=3.1\times$10$^{45}$ erg/s in the 2-10 keV range,  \citet{Allen1998})
cluster Abell~2744 (also known as AC118, or RXCJ0014.3-3022) 
at a redshift of  $z=0.308$ \citep{Couch1984}. 
The $\sim 250$ kpc offset is larger than that in the \textit{Bullet Cluster} -- although, as we shall discuss later,
this value does not describe the separation of the main mass-clump from its stripped gas component. 
Nevertheless, Abell~2744 is undergoing a particularly interesting merger. 
The complex interplay between multiple dark matter and baryonic components appears to have unleashed `ghost', `dark', `stripped' and `bullet' clusters.

The first hint that Abell~2744 is in the middle of a major merging event arose from observation of a powerful and extended radio halo 
($P(1.4 \textrm{GHz}) > 1.6\times 10^{36}$ Watt,~
\citet{Giovannini1999,Govoni2001,Govoni2001a}). 
This indicated the presence of relativistic electrons accelerated through 
high Mach shocks or turbulence \citep[e.g.][]{Sarazin2004}.
The picture was clarified by X-ray studies \citep{Kempner2004,Zhang2004} that
revealed substructure near the cluster core, plus an additional luminous structure towards the Northwest. 
Kinematic observations of cluster member galaxies \citep{Girardi2001}
suggested a bimodal distribution in redshift space, but were not at first considered significant.  
Recent kinematic studies focussing solely on Abell~2744 definitely show a bimodal velocity dispersion 
in the cluster centre, together with a third group of cluster members near the Northwestern X-ray peak \citep{Boschin2006,Braglia2009}.
Although there is no evidence for non-thermal X-ray emission \citep{Million2009}, the fraction of 
blue star-forming galaxies \citep[][and references therein]{Braglia2007} also seems to be enhanced.
A default explanation emerged for the centre of Abell~2744, featuring a major merger in the North-South direction with a small
inclination towards the line-of-sight and a $\sim$~3:1 mass ratio of the merging entities \citep{Kempner2004,Boschin2006}.
More controversial is the role of the Northwestern structure.
\citet{Kempner2004} detected a cold front on its SW edge and a possible shock front 
towards the cluster core, so concluded that it is falling towards the main mass. 
More recent analysis of  {\sl Chandra} data \citep{Owers2011} failed to confirm the presence of the shock 
front, only a cold front towards the Northern edge, and the authors proposed that the structure is moving 
instead towards the North/Northeast, after being deﬂected from the main cluster in an off-centre core passage.
What seems sure is that we are at least observing a complicated merger between three separate bodies \citep{Braglia2007}.

Until now, Abell~2744 has been
better constrained from X-ray and kinematic studies than by gravitational lensing.
So far, \citet{Smail1997} detected a weak-lensing signal and strong-lensing features, followed by \citet{Allen1998} who
found a large discrepancy in the mass estimates for Abell~2744 from X-ray and strong-lensing reconstructions. Given the cluster's dynamical
state, this finding is perhaps no longer surprising since the merging would induce non-thermal support and elongation along the line-of-sight, 
which increases the systematics from both methods. 
The most recent weak-lensing analysis \citep{Cypriano2004} showed indications of substructure
in the reconstructed surface-mass density, but did not reach the resolution required for more quantitative statements. 
Here, we present the results of an {\sl HST} imaging survey, aimed at clarifying the 
evolutionary stage of this complex system.

This article is organised as follows. In Sec.~\ref{LENSING} we present our comprehensive lensing analysis, mainly based on
newly acquired data taken with the {\sl Advanced Camera for Surveys (ACS)} on the {\sl Hubble Space Telescope (HST)}. 
In Sec.~\ref{XRAY} we describe a complementary X-ray analysis based on {\sl Chandra} data.
We discuss cosmological implications and an interpretation of the merging scenario 
in Sec.~\ref{INTERPRETATION}, and we conclude in Sec.~\ref{CONCLUSIONS}. Throughout this paper we assume a cosmological 
model with $\Omega_{\textrm{m}}=0.3$, $\Omega_{\Lambda}=0.7$ and $\textrm{h}=0.7$. 
At the cluster's redshift $z=0.308$, one arcsecond corresponds to $4.536$~kpc.


\section{Lensing analysis}
\label{LENSING}
Our recently acquired, multiband {\sl HST/ACS} imaging enables us to significantly improve upon previous mass models of Abell~2744.  
Using the parametric method described in Sec.~\ref{SLENSING}, we have identified strong gravitational
lensing of 11 background galaxies producing 34 multiple images around the Southern core,
with an Einstein radius of $r_{\textrm{E}} \sim 30\arcsec$ (see below).
These enable us to tightly constrain the position and shape of the core mass distribution.
No such multiple image systems are revealed around the N or NW clumps, immediately indicating that their masses are lower.
Our {\sl HST} images also yield $\sim62~\textrm{galaxies}/\textrm{arcmin}^{2}$ for weak lensing analysis 
(after charge transfer inefficiency (CTI) corrections are performed, as described below).  This enables detailed mass modelling throughout our {\sl HST} field of view.  
To probe the cluster merger on even larger scales, we incorporate ground-based weak lensing measurements from {\sl VLT} and {\sl Subaru}.

\begin{figure*}
\begin{center}
 \includegraphics[width=.75\textwidth]{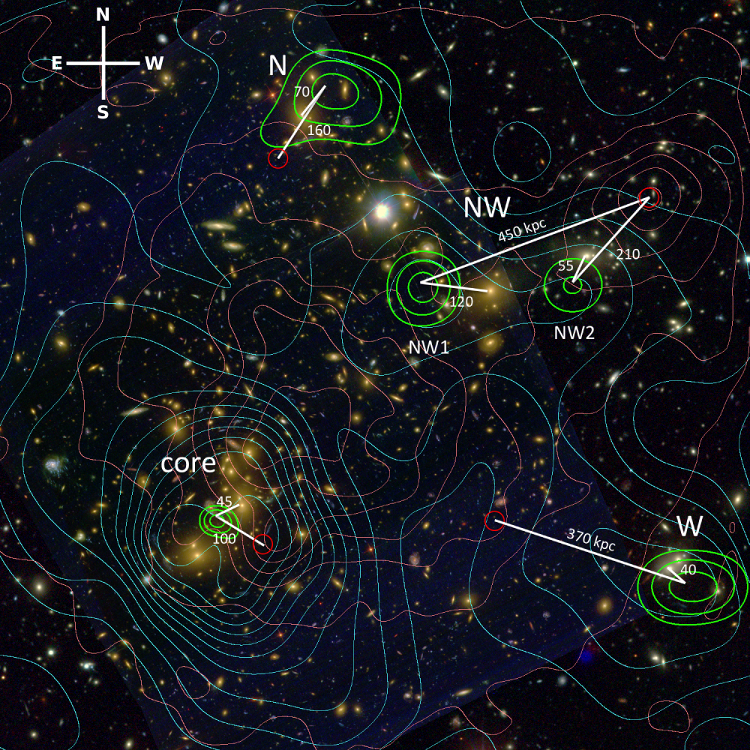}
\end{center}
\caption{The field of Abell~2744, with different interesting features. The false-colour background is provided by {\sl HST/ACS}
(the two pointings can be identified by the higher, bluely background noise level), {\sl VLT} and {\sl Subaru} images on a field
 size of $240\arcsec\times 240\arcsec$ ($\sim 1.1$ Mpc on a side). Overlaid in cyan are the surface-mass density contours most concentrated in the 'core' area
 and in magenta the more evenly-distributed X-ray luminosity contours. 
The peak positions of the core, N, NW, and W clumps are indicated by the green likelihood contours, derived from the bootstrap samples.  
Contours are 86\%, 61\%, and 37\% of the peak likelihood for each clump. (Assuming a Gaussian probability distribution, 
these would correspond to 0.3, 1, and 2-sigma confidence contours.) Note that the NW clump most likely peaks in the area 
indicated by NW1 (in 95\% of the bootstrap realisations), but peaks also at NW2 in 54\% of the bootstrap realisations.
The small red circles show the position of the local overdensities in the gas distribution, 
associated with each individual dark matter clump. The white rulers show the separation between dark matter peaks and the bright 
clusters galaxies and local gas peaks.}
\label{DMXOVERLAY}
\end{figure*}

We simultaneously fit all of these strong- and weak-lensing observations using our well-tested mass reconstruction algorithm \citep{Merten2009,Meneghetti2010a} 
(which is similar to that used by \citet{Bradav2006} and \citet{Bradav2008} to map the \textit{Bullet Cluster}
and the \textit{Baby Bullet}).  Importantly, we make no assumptions about mass tracing light in the combined analysis. It should
be noted though, that we rely on a parametric method, assuming that light traces mass, to identify multiple image systems. We account
for possible uncertainties in this identification, by resampling the set of less confident identifications in the errors analysis based on a 
bootstrapping technique (see Sec. \ref{WLSL}). 
Our analysis reveals {\em four} individual clumps of mass $\simgt 10^{14} M_{\odot}$ within a $250$ kpc radius.  
Previous weak lensing analysis of {\sl VLT} images alone had resolved but a single broad mass clump \citep{Cypriano2004}. 
Below we describe our datasets, analyses, and results in more detail. The central cluster field and a preview on the matter and gas
distribution is presented in Fig.~\ref{DMXOVERLAY}, which also shows the contours encircling different areas of 
specific probability, that a mass peak is located at this specific location of the reconstructed field. 

\subsection{The HST/ACS dataset}
\label{ACSdata}
The {\sl HST} data consist of two pointings in Cycle 17 (data taken between Oct.~27-30 2009,
Proposal~ID:~11689, P.I.:~R.~Dupke) 
with $\sim 50\%$ overlap between the pointings. The images were taken
with the {\sl ACS/WFC} camera using three different filters, F435W ($16.2~\textrm{ksec}$\footnote{Equally split between the two pointings}), 
F606W ($13.3~\textrm{ksec}$) and F814W ($13.2~\textrm{ksec}$).

The {\sl HST/ACS} camera had been in orbit for eight years when the imaging was acquired. 
During this time above the protection of the Earth's atmosphere, its CCD detectors had been 
irreparably damaged by a bombardment of high energy particles.
During CCD readout, photoelectrons are transported to the readout amplifier through a silicon lattice.
Damage to this lattice creates charge traps that delay some electrons and spuriously trail the image -- in a way that
alters the shapes of galaxies more than the gravitational lensing signal that we are trying to measure. 
To undo this trailing and correct the {\tt \_raw} images pixel-by-pixel, we used the detector readout model of \citet{Massey2010}, 
updated for device performance post Servicing Mission 4 by \citet{Massey2010a}. The corrected data was then reduced via the standard CALACS 
pipeline \citep{Pavlovsky2006}, and stacked using multidrizzle \citep{Koekemoer2002}.

\subsection{Strong lensing}
\label{SLENSING}
\begin{figure}
\begin{center}
 \includegraphics[width=.45\textwidth]{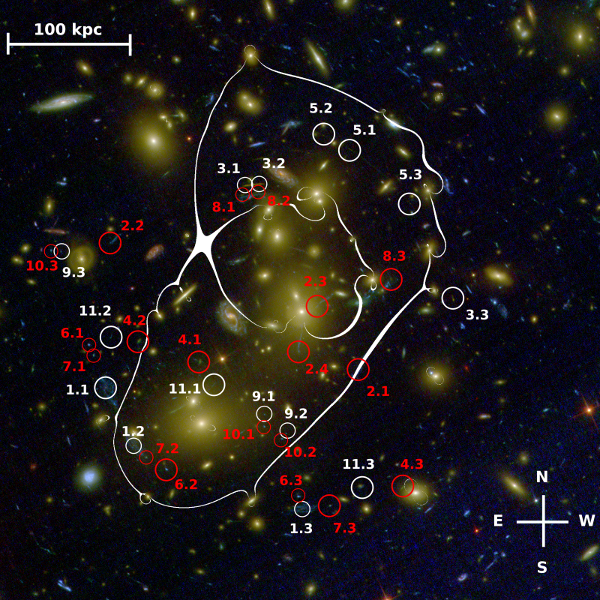}
 \end{center} 
\caption{A zoom into the innermost core region of the {\sl HST/ACS} images. 
Shown as a continuous white line is the critical curve of the cluster as it is derived from the strong-lensing model. 
It assumes a source redshift $z_{\textrm{s}}=2.0$. Also shown are the approximate positions of the 
identified multiple-image systems as they are listed in Tab.~\ref{MSYSTEMS} with a varying two-colour scheme to avoid 
confusion of close-by images. These systems were found by the method of
\citet{Zitrin2009} and are not simple optical identifications.
The visible field size is $\sim100\arcsec\times100\arcsec$, translating to $\sim~450$ kpc on a side.}
 \label{ACS_RGB}
\end{figure}

We concentrate our strong-lensing analysis on the reduced {\sl ACS} images only.
Several strong-lensing features are immediately identifiable by eye on the combined three-band image (Fig.~\ref{ACS_RGB}).  
To find additional multiple images across the field of view, we apply the well-tested approach of \citet{Zitrin2009} \citep[see also][]{Zitrin2011b} to 
lens modelling, which has previously uncovered large numbers of multiply-lensed galaxies in {\sl ACS}
images of Abell 1689, Cl0024, 12 high-$z$ MACS clusters, MS1358 and Abell 383 \citep[respectively,][]{Broadhurst2005,
Zitrin2009,Zitrin2011,Zitrin2010a,Zitrin2011a,Zitrin2011c} .

In the \citet{Zitrin2009} method, the large-scale distribution of cluster mass is approximated by assigning a
 power-law mass profile to each galaxy, the
 sum of which is then smoothed. The
 degree of smoothing ($S$) and the index of the power-law ($q$) are
 the most important free parameters determining the mass profile. A
 worthwhile improvement in fitting the location of the lensed images
 is generally found by expanding to first order the gravitational
 potential of this smooth component, equivalent to a coherent shear
 describing the overall matter ellipticity, where the direction of the
 shear and its amplitude are free parameters. This allows for some flexibility in
 the relation between the distribution of dark matter and the distribution of
 galaxies, which cannot be expected to trace each other in detail. The
 total deflection field $\vec{\alpha}_T(\vec\theta)$, consists of the
 galaxy component, $\vec{\alpha}_{gal}(\vec\theta)$, scaled by a
 factor $K_{gal}$, the cluster dark matter component
 $\vec\alpha_{DM}(\vec\theta)$, scaled by (1-$K_{gal}$), and the
 external shear component $\vec\alpha_{ex}(\vec\theta)$:
\begin{equation}
\label{defTotAdd}
\vec\alpha_T(\vec\theta)= K_{gal} \vec{\alpha}_{gal}(\vec\theta)
+(1-K_{gal}) \vec\alpha_{DM}(\vec\theta)
+\vec\alpha_{ex}(\vec\theta),
\end{equation}
where the deflection field at position $\vec\theta_m$
due to the external shear,
$\vec{\alpha}_{ex}(\vec\theta_m)=(\alpha_{ex,x},\alpha_{ex,y})$,
is given by:
\begin{equation}
\label{shearsx}
\alpha_{ex,x}(\vec\theta_m)
= |\gamma| \cos(2\phi_{\gamma})\Delta x_m
+ |\gamma| \sin(2\phi_{\gamma})\Delta y_m,
\end{equation}
\begin{equation}
\label{shearsy}
\alpha_{ex,y}(\vec\theta_m)
= |\gamma| \sin(2\phi_{\gamma})\Delta x_m -
  |\gamma| \cos(2\phi_{\gamma})\Delta y_m,
\end{equation}
where $(\Delta x_m,\Delta y_m)$ is the displacement vector of the
position $\vec\theta_m$ with respect to a fiducial reference position,
which we take as the lower-left pixel position $(1,1)$, and
$\phi_{\gamma}$ is the position angle of the spin-2 external
gravitational shear measured anti-clockwise from the $x$-axis.  The
normalisation of the model and the relative scaling of the smooth dark matter
component versus the galaxy contribution brings the total number of
free parameters in the model to 6. This approach to strong lensing is sufficient
to accurately predict the locations and internal structure of multiple
images, since in practice the number of multiple images
readily exceeds the number of free parameters, which become fully constrained.

Two of the 6 free parameters, namely the galaxy power law index $q$
and the smoothing degree $S$, can be initially set to
reasonable values so that only 4 of the free parameters need to be fitted at first.
This sets a very reliable starting-point using obvious
systems. The mass distribution is therefore well-constrained and
uncovers many multiple-images that can be
iteratively incorporated into the model, by using their redshift
estimation, from photometry, spectroscopy or model prediction and location in the image-plane.
At each stage of the iteration, we use the model to lens the most obvious lensed
galaxies back to the source plane by subtracting the derived
deflection field, then relens the source plane to predict the
detailed appearance and location of additional counter images, which
may then be identified in the data by morphology, internal structure
and colour. We stress that multiple images found this way must be
accurately reproduced by our model and are not simply eyeball
``candidates'' requiring redshift verification. The best fit is
assessed by the minimum RMS uncertainty in the image plane
\begin{equation} \label{RMS}
RMS_{\textrm{images}}^{2}=\sum_{i} ((x_{i}^{'}-x_{i})^2 + (y_{i}^{'}-y_{i})^2) ~/ ~N_{\textrm{images}},
\end{equation}
where $x_{i}^{'}$ and $y_{i}^{'}$ are the locations given by the
model, $x_{i}$ and $y_{i}$ are the real image locations, and the
sum is over the total number of images $N_{\textrm{images}}$. The best-fit solution is unique
in this context, and the model uncertainty is determined by the
locations of predicted images in the image plane. Importantly, this
image-plane minimisation does not suffer from the well known bias
involved with source plane minimisation, where solutions are biased by
minimal scatter towards shallow mass profiles with correspondingly
higher magnification.

In Abell 2744 we have uncovered a total number of 34 multiple images belonging to 11 background sources. 
We label the different systems in Fig.~\ref{ACS_RGB}, which also shows the critical curve of the cluster derived from
the strong lensing model. The model predicts an Einstein radius of $r_{\textrm{E}} = 30\arcsec\pm3\arcsec$. 
With such a large number of clear multiple images we can constrain the inner 
mass distribution very well, and we shall incorporate these constraints into our joint strong- and weak-lensing analysis described in Sec.~\ref{WLSL}.

\begin{table}
 \caption{The multiple-image system of Abell~2744.}
 \label{MSYSTEMS}
\begin{center}
 \begin{tabular}{cccc}
  \hline
  Image-ID & $x$ & $y$& $z$ \\
($\langle\textrm{\tiny{source}}\rangle$.$\langle\textrm{\tiny{image}}\rangle\langle\textrm{\tiny{additional knot}}\rangle$)&(\arcsec)&(\arcsec)&\\
  \hline
   1.1&$-35.10$&  $-13.55$&$2.0\pm0.3$\\
   1.2&$-30.15$&  $-23.95$&\\
   1.3&$0.10$&  $-35.40$&\\
  1.11&  $-33.60$&  $-16.50$&\\
  1.21&  $-31.50$&  $-21.55$&\\
  1.31&    $1.60$&  $-35.85$&\\
\hline
   2.1&    $9.30$&  $-11.50$&$2.0\pm0.3$\\
   2.2&  $-34.25$&   $12.35$&\\
   2.3&    $2.80$&    $1.05$&\\
   (2.4)&   $-0.50$&   $-7.10$&\\
  2.11&   $11.55$&   $-7.65$&\\
  2.21&  $-32.55$&   $13.90$&\\
  2.31&    $5.55$&    $3.10$&\\
  (2.41)&   $-0.05$&   $-4.15$&\\
\hline
   3.1&  $-10.10$&   $22.65$&$4.0\pm0.3$\\
   3.2&   $-7.40$&   $22.95$&\\
   (3.3)&   $27.15$&    $2.20$&\\
\hline
   4.1&  $-18.25$&   $-9.00$&$3.5\pm0.3$\\
   4.2&  $-29.25$&   $-5.30$&\\
   4.3&   $18.05$&  $-31.60$&\\
\hline
   5.1$^{*}$&    $8.85$&   $29.00$&$4.0\pm0.5$\\
   5.2$^{*}$&    $3.85$&   $31.60$&\\
   5.3$^{*}$&   $19.65$&   $19.30$&\\
\hline
   6.1$^{*}$&  $-38.15$&   $-5.95$&$3.0\pm0.5$\\
   6.2$^{*}$&  $-24.25$&  $-28.30$&\\
   6.3$^{*}$&   $-0.60$&  $-33.15$&\\
\hline
   7.1$^{*}$&  $-37.35$&   $-7.85$&$3.7\pm0.5$\\
   7.2$^{*}$&  $-27.85$&  $-26.10$&\\
   7.3$^{*}$&    $5.10$&  $-34.85$&\\
\hline
   8.1&  $-10.70$&   $20.90$&$4.0\pm0.2$\\
   8.2&   $-8.00$&   $21.40$&\\
   (8.3)&   $16.10$&    $5.90$&\\
\hline
   9.1$^{*}$&   $-6.65$&  $-18.45$&$3.0\pm0.5$\\
   9.2$^{*}$&   $-2.80$&  $-21.90$&\\
   (9.3)&  $-43.20$&   $10.80$&\\
\hline
  10.1$^{*}$&   $-6.65$&  $-20.65$&$3.0\pm0.5$\\
  10.2$^{*}$&   $-3.55$&  $-22.75$&\\
  10.3$^{*}$&  $-44.90$&   $11.00$&\\
\hline
  (11.1)&  $-16.00$&  $-13.30$&$3.0\pm0.5$\\
  (11.2)&  $-34.20$&   $-4.65$&\\
  (11.3)&   $10.70$&  $-31.55$&\\
\hline
 \end{tabular}
\end{center}
\medskip
If the image ID is shown in brackets, the image is not confidently reproduced by the lensing model described in Sec.~\ref{SLENSING}.
The systems marked with an asterisk, were nicely reproduced by the parametric lensing model but we allow them to be resampled
by our nonparametric reconstruction technique to allow for uncertainties in the number of identified strong lensing features.
All other images define a `confident catalogue' of multiple-image systems and are used in our subsequent analysis.
The x-and y-coordinates are relative to the BCG position ($\alpha_{\textrm{J2000}} =3.58611^{\circ}$, $\delta_{\textrm{J2000}} =-30.40024^{\circ}$)
in arcseconds. Redshifts of each system and their respective error are derived from the model predictions.
\end{table}

\subsection{Weak lensing}
\label{WLENSING}
Several areas outside the cluster core are of special interest, due to the complicated structure of this merging cluster. 
We measure the shapes of background galaxies to derive a weak-lensing signal and extend our mass reconstruction over a much larger field.  
Since our {\sl HST} data cover only a limited field of view, we also include {\sl VLT} and {\sl Subaru} imaging in our weak lensing analysis.
Additional {\sl HST} pointings in the future are crucial for a full interpretation of this merger, as we shall discuss in the course of
this analysis.

\subsubsection{HST/ACS}
To select lensed background galaxies, we obtained photometric redshifts 
based on our three filters using Bayesian photometric redshifts (BPZ) \citep{Ben'itez2000,Coe2006}.
Cluster ellipticals at $z \sim 0.3$ occupy a unique region in our colour-colour space,
enabling us to obtain better than expected results.

Of 118 galaxies with published spectroscopic redshifts 
(Owers et al. 2011 and references therein; all $z < 0.7$) within our field of view, 
99 yield confident photo-zs, accurate to $\Delta z \sim 0.06(1+z)$ RMS with no significant outliers.
Background galaxies were selected as those with confident photo-z $> 0.5$.
This cut successfully excludes all 5 foreground and 86 cluster galaxies in our spec-z sample with confident photo-z 
(though one cluster galaxy was assigned a less confident photo-z $\sim$ 0.8).  
However, given our limited photometric coverage, we do not rule out some contamination of our background 
sample with foreground and/or cluster galaxies.

We measure the weak gravitational lensing signal in the F814W {\sl HST} exposures, 
which are the deepest and contain the most galaxies at high redshift. 
We measure their shapes, and correct them for convolution by the Point-Spread Function (PSF), using the `RRG'  \citep{Rhodes2000} pipeline
developed for the {\sl HST} COSMOS survey \citep{Leauthaud2007,Massey2007b}.
The RRG method is particularly optimised for use on high resolution, space-based data.
Since {\sl HST} expands and contracts as it warms in the sun or passes through the shadow of the Earth, even this telescope does not have a constant PSF.
However, 97\% of the variation in its PSF can be accounted for by variation in its focal length \citep[][the separation between the primary and secondary mirrors]{Jee2007a}.
We therefore measure its focal length by matching the shapes of the $\sim 12$ bright stars in each pointing to models created by raytracing through the optical design \citep{Krist2003}. 
This achieves a repeatable precision of $1\mu m$ in the determination of the focal length, and we construct a PSF model from all those stars observed during the 600-orbit COSMOS survey at a similar focus \citep{Rhodes2007}.
We finally use this PSF model to correct the shapes of galaxies, and to obtain estimates of the amount by which their light has been sheared. 
The result is a catalogue of $1205$ galaxies with shear estimates, corresponding to a density of $\sim62~\textrm{galaxies}/\textrm{arcmin}^{2}$.

\subsubsection{VLT/FORS1}
The complementary {\sl VLT} data for our weak lensing analysis is identical to that included by \citet{Cypriano2004} in a study of 24 X-ray Abell clusters. 
The total field of view is $6\farcm8$ on a side, centred on the brightest cluster galaxy (BCG) and significantly exceeding the coverage of our {\sl HST} imaging.
$V$, $R$ and $I$-band imaging was obtained with the {\sl FORS1} camera between April and July 2001, with exposure times of $330$ s in each filter.
The data were reduced with standard IRAF routines and, to maximise the depth, we perform weak-lensing shape analysis on the combined VRI image.
Seeing conditions were excellent, with a FWHM of stars in the combined VRI image of $0\farcs59$.

We measure galaxy shapes and perform PSF correction using the \texttt{IM2SHAPE} method \citep{Bridle2002}.
This involves a two-step process to first map the PSF variation across the observed field using stars, then to model each detected galaxy, perform PSF correction and recover its ellipticity. 
To remove foreground contamination and unreliable shape measurements from our catalogue, we apply magnitude cuts plus additional rejection criteria \citep[see][]{Cypriano2004}. 
Since that work, we have improved the efficiency of foreground galaxy removal and now keep a higher density of background galaxies in our shear catalogue.
The result is a catalogue of $912$ galaxies with shear estimates, or $\sim 20~\textrm{galaxies}/\textrm{arcmin}^{2}$.

\subsubsection{Subaru/SuprimeCam}
To extend the total field of view even further, especially in the Northern areas of the cluster field, we obtained $1.68$ ksec $i'$-band imaging data with {\sl Subaru/SuprimeCam} during Semester S08B. 
The data were reduced following \citet{Okabe2008} and \citet{Okabe2010,Okabe2010a}.
Astrometric calibration was conducted by fitting the final stacked image with the 2MASS data point source catalogue; residual astrometric errors were less than the CCD pixel size.
Due to poor weather conditions, the seeing size is as large as $1\farcs28$.

We measure galaxy shapes and perform PSF correction using the \texttt{IMCAT} package \citep[provided by][]{Kaiser1995} in the same pipeline as \citet{Okabe2010,Okabe2010a} with some modifications following \citet{Erben2001}. 
Background galaxies were selected in the range of $22<i'<26~{\rm ABmag}$ and
${\bar r}_h^*+\sigma_{r_h^*}\simeq3.4<r_h<6.0~{\rm pix}$,  
where $r_h$ is the half-light radius, and ${\bar r}_h^*$ and
$\sigma_{r_h^*}$ are the median and standard error of
stellar half-light radii $r_{h^*}$, corresponding to the half median
width of the circularised PSF. 
The density of background galaxies in our final shear catalogue is $\sim 15~\textrm{galaxies}/\textrm{arcmin}^{2}$.
This is 30-50\% of typical values from images obtained during normal weather conditions \citet{Okabe2010,Okabe2010a}.

\subsection{Combined lensing reconstruction}
\label{WLSL}

In order to combine the weak- and strong-lensing constraints in a consistent way, we use the joint lensing reconstruction algorithm
described in \citet{Merten2009} \citep[see also][for a similar approach]{Bradav2005,Bradav2009}. This method has been extensively 
tested in \citet{Meneghetti2010a} and proved its capability to faithfully recover the cluster mass distribution over a broad range of scales.

Our joint mass reconstruction is nonparametric, in the sense that it neither makes any \textit{a priori} assumptions
about the cluster's underlying mass distribution nor does it need to trace any light-emitting component in the
observed field. However, as described in Sec.~\ref{SLENSING}, we use a parametric method to identify the multiple image systems 
in the field.
We reconstruct the cluster's lensing potential (its gravitational potential projected onto the plane of the sky) $\psi$ by combining measurements
of the position of the critical line and the reduced shear.
To do this, we divide the observed field into an adaptive mesh, which discretises all observed and reconstructed quantities. 
A statistical approach is chosen to combine our various measurements,
by defining a multi-component $\chi^{2}$-function that depends on the underlying lensing potential and a regularisation term
$R(\psi)$ to prevent the reconstruction overfitting noise \citep[see][]{Merten2009,Bradav2005}
\begin{equation}
 \chi^{2}(\psi) = \chi^{2}_{\textrm{w}}(\psi)+\chi^{2}_{\textrm{s}}(\psi)+R(\psi).
\end{equation}

The weak-lensing term is defined by the expectation value of the complex reduced shear in each mesh position $\left\langle\varepsilon\right\rangle$, which is obtained
by averaging the measured ellipticities of all background galaxies within that grid cell
\begin{equation}
  \chi^{2}_{\rm w}(\psi)=\left(
   \left\langle\epsilon\right\rangle -\frac{Z(z)\gamma(\psi)}{1-Z(z)\kappa(\psi)}
 \right)_{i}\mathcal{C}^{-1}_{ij}\left(
   \left\langle\epsilon\right\rangle -\frac{Z(z)\gamma(\psi)}{1-Z(z)\kappa(\psi)}
 \right)_{j} ,
\end{equation}
where $Z(z)$ is a cosmological weight factor as defined e.g.~in \citet{Bartelmann2001}, and $\gamma = \partial\partial\psi/2$ 
is the shear of the lens, with the two components expressed in complex notation. 
$\kappa = \partial\partial^{*}\psi$ is the convergence, where the complex differential operator in
the plane is defined as $\partial := (\frac{\partial}{\partial\theta_{1}} +\mathsf{i}\frac{\partial}{\partial\theta_{2}})$, with $\theta_{1}$ 
and $\theta_{2}$ being the two angular coordinates in the sky. The indices $i,j$ indicate the discretisation of the input data and
the lens properties, where we have to take into account the full $\chi^{2}$-function because the averaging process of background
galaxies might result in an overlap of neighbouring mesh points, expressed by the covariance matrix $\mathcal{C}_{ij}$.

The strong-lensing term is defined as 
\begin{equation}
 \chi^{2}_{{\rm s}}(\psi)=\frac{\left(\det{\mathcal A}(\psi)\right)^{2}_{k}}{\sigma^{2}_{{\rm s}}}=
 \frac{\left((1-Z(z)\kappa(\psi))^{2}-|Z(z)\gamma(\psi)|^{2}\right)^{2}_{k}}{\sigma^{2}_{{\rm s}}}\;,
\end{equation}
where the index $k$ labels all pixels in the reconstruction mesh, which are supposed to be part
of the critical curve within the uncertainties $\sigma_{\rm s}$, given by the pixel size of the grid. 
At these points, the Jacobian determinant $\det{\mathcal A}(\psi)$ of the lens mapping must vanish.

We iterate towards a best-fitting lens potential by minimising the $\chi^{2}$-function at each mesh position
\begin{equation}
  \frac{\partial\chi^{2}(\psi)}{\partial\psi_{l}}\stackrel{\textrm{!}}{=}0 \quad \mathsf{with} \quad l\in[0,...,N_{\mathsf{pix}}]\;.
\end{equation}
In practice, we achieve this by translating this operation into a linear system of equations and invoking a two-level iteration
scheme \citep[see][and references therein]{Merten2009}.

In this analysis, we use all strongly lensed multiple-image systems from the confident sample (Tab.~\ref{MSYSTEMS}) together with their derived redshifts, and 
combine weak lensing shear catalogues from all three telescopes (see Sec.~\ref{WLENSING}) as described in Sec.~\ref{CMERGING}.
The combination of strong lensing data and the density of background galaxies allows for a reconstruction
on a mesh of $72\times72$ pixels in the central region (corresponding to a pixel-scale of $8.4\arcsec/\textrm{pix}$)
and $36\times36$ pixels in the outskirts of the field (corresponding to a pixel-scale of $16.7\arcsec/\textrm{pix}$).
Error estimates were produced by bootstrapping the redshift uncertainties of the strong-lensing constraints, and 
resampling those multiple image systems in Tab.~\ref{MSYSTEMS}, which are marked with an asterisk. The result are 500 different bootstrap
realisations of the refined cluster core. 150 bootstrap realisations of the cluster outskirts were produced by 
bootstrapping the combined ellipticity catalogues. The number of bootstrap realisations is mainly constrained by runtime considerations. All error estimates 
are calculated from the scatter within the full bootstrap sample. If not stated differently, the given value reflects 
68\% confidence level.

\subsubsection{Combining the catalogues}
\label{CMERGING}
In order to combine the three different catalogues of ellipticity measurements we used the following strategy. Clearly, the {\sl Subaru}
catalogue delivered the largest field-of-view but it was derived from only single-band imaging and under bad seeing conditions. Therefore,
we decided to limit its field size to a box of $600\arcsec \times 600\arcsec$ around the centre of the cluster and to cut out the central
$\sim 400\arcsec \times 400\arcsec$ part, which was sufficiently covered by {\sl HST/ACS} and {\sl VLT} exposures with better data quality.
As a result, {\sl Subaru} data only covers the outermost 200\arcsec on each side of the field.\\
One might argue that the combination of  {\sl HST/ACS} and {\sl VLT} ellipticities in the innermost centre of the cluster is problematic,
but indeed the galaxy density is three times higher in the {\sl HST/ACS} field and clearly dominating the adaptive-averaging process
of the nonparametric reconstruction algorithm that we used for the further analysis (compare Sec.~\ref{WLSL} and \citet{Merten2009}).\\
Another issue is the possible double-counting of galaxies, so we cross-correlated both catalogues with a correlation radius of 2.5\arcsec
and found 160 double-count candidates over the full {\sl HST/ACS} field. Given the pixel size of the final reconstruction, this translates
to an insignificant average double-count probability of less than one galaxy per pixel. 
Furthermore, the weighting scheme of the adaptive-averaging process was implemented such, that the highest weight of the {\sl VLT} ellipticities
was identical to the smallest weight of the {\sl HST/ACS} ellipticities. Finally, the errors in the ellipticity measurement 
are treated in the joint reconstruction method in a purely statistical way by deriving the variance of the weight-averaged sample 
of ellipticities in each reconstruction pixel.\\
To derive a physical surface mass density from the scaled lensing convergence one needs to know at least the mean redshift of the background
galaxy population that was used to produce the ellipticity catalogues.   
Problems with the different depths of the fields and therefore
with the final mass analysis should not be a crucial issue for a relatively low-redshift cluster like Abell~2744. However, the redshift for
determining the surface-mass density from the reconstructed convergence in the overlap area has been calculated as the galaxy-density weighted average of 
both, the {\sl VLT} and the {\sl HST/ACS} populations. The redshifts of the strong lensing features (compare Tab.~\ref{MSYSTEMS})
were included for the determination of the core mass and the respective redshifts of the {\sl VLT} and {\sl Subaru} source distribution
was used in the outskirts of the field.

In order to test the effect of dilution in the outskirts of the field, we increased
the ellipticity values for all Subaru background galaxies by 15\% and repeated the reconstruction. This test is necessary due to
the single-band {\sl Subaru} imaging. As it turns out, the difference in the reconstructed convergence is marginal since the
ellipticity values are already low with large scatter in this area of the field. However, the effect was included in the determination 
of the error budget for the reconstructed total mass and mass profile.

\subsection{Reconstruction results}
\label{LENSING_REC}

\begin{figure}
 \begin{center}
  \includegraphics[width = .48\textwidth]{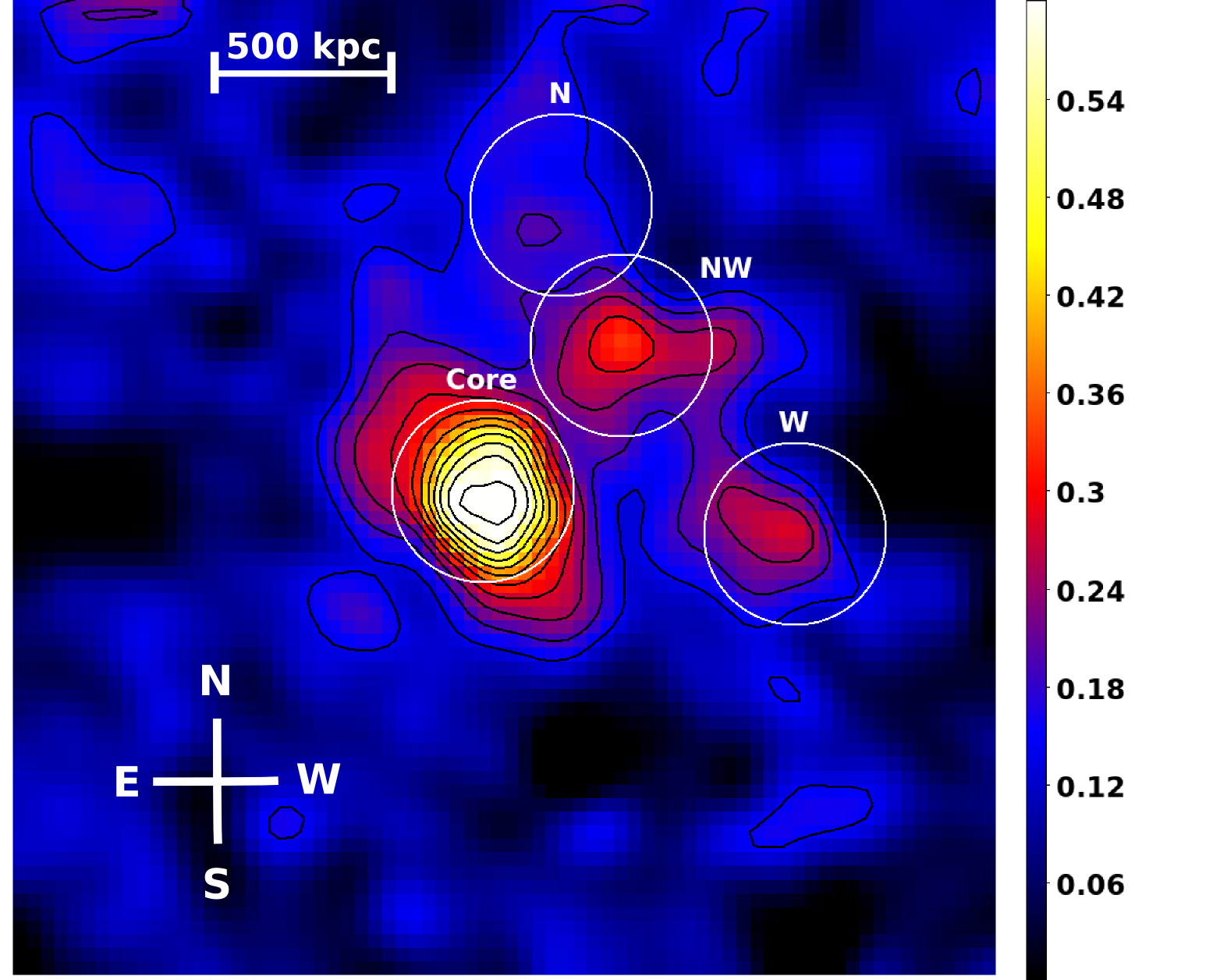}
 \end{center}
\caption{The convergence map of the cluster field for a source redshift of $z_{\textrm{s}}=\infty$ and a field size of $600\arcsec\times600\arcsec$, translating to $\sim 2.7$ Mpc on a side.
The black contours start at $\kappa_{0}=0.14$ with a linear spacing of $\Delta\kappa=0.047$. The four white circles with labels indicate
identified sub-clumps and the radius within which their mass is calculated. The radius of all four circles is $55.4\arcsec\approx250$~kpc.}
\label{CONV}
\end{figure}

\begin{figure}
 \begin{center}
  \includegraphics[width = .5\textwidth]{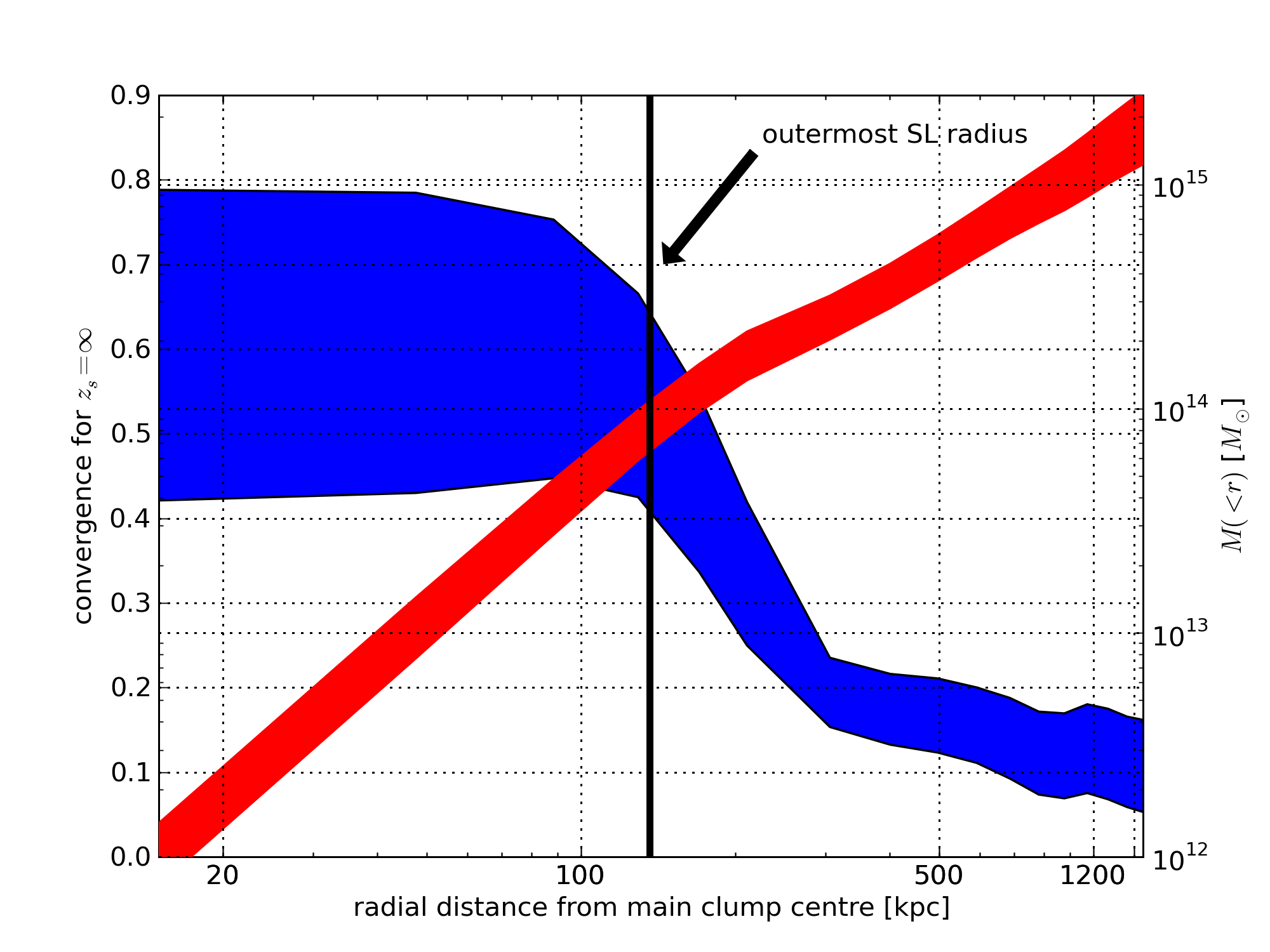}
 \end{center}
\caption{Radial mass profiles of the cluster, derived from our lensing analysis. Shown in blue (referring to the left $y$-axis) is the convergence profile for a source redshift $z_{s}=\infty$.
The large uncertainty for small radii arises from uncertainty in the exact position of the cluster centre. The cumulative mass
as a function of radial distance from the cluster centre is shown in red (referring to the right $y$-axis). The black vertical line indicates the distance of the 
outermost multiple images from the cluster centre.} 
\label{RPROFILE}
\end{figure}

We obtain a map of the lensing convergence across the field (proportional to the projected mass) 
by applying the Laplacian operator to the lensing potential on the adaptively refined mesh (see Fig.~\ref{CONV}).
We find a total mass (by assuming $h=0.7$ here and further on) within a radius of $1.3$~Mpc around the Core of $M(r<1.3\textrm{Mpc})=1.8\pm0.4\times 10^{15}M_{\odot}$, which is
in good agreement with kinematically derived masses \citep{Boschin2006}. 
A mass determination within a field of ($1300 \times 750$) kpc (to compare easily to the work on the {\it Bullet Cluster} of \citet{Bradav2006}) 
centred on the Core density peak 
yields $M(1.73 \textrm{Mpc}^{2}) = 7.4 \pm 1.0\times 10^{14} M_{\odot}$, rendering Abell~2744 comparable in mass or slightly less massive 
than the \textit{Bullet Cluster} \citep{Bradav2006}.
The overall radial convergence and mass profile can be found in Fig.~\ref{RPROFILE}.

Most interestingly, our gravitational lensing analysis resolves four distinct sub-structures, indicated in Fig.~\ref{CONV} by the white circles. 
We label these substructures as Core, Northwestern (NW, later on dubbed as `dark'), Western (W, later on dubbed as `stripped') and Northern (N) structure.
Please note that there is some indication of a possible double peak in the NW structure, as can be seen in Fig.~\ref{DMXOVERLAY}
and also in the clearly visible extension of the structure towards the West in Fig.~\ref{CONV}.
The Core, NW and W clumps are clear detections in the surface-mass density distribution with $11\sigma$, $4.9\sigma$ and $3.8\sigma$ 
significance over the background level, respectively. 
\footnote{The background level was estimated in the following way: From the total convergence, shown in Fig.~\ref{CONV}, a radial area 
of $75\arcsec$ around the Core, $70\arcsec$ around NW and $55\arcsec$ around N and W was cut out. From the remaining field the
mean and variance in the convergence level was calculated.}
Somewhat fainter with $2.3\sigma$ significance is the N structure, but it clearly
coincides with a prominent X-ray substructure found by \citet{Owers2011} and is therefore included in the further analysis (compare Sec.~\ref{XRAY}).  
The positions of the mass peaks and their local projected masses within $250$~kpc are listed in Tab.~\ref{STRUCTURES} and shown in
more detail in Fig.~\ref{DMXOVERLAY}.
The new {\sl HST/ACS} images thus allow a striking improvement in our map of the mass distribution and reveal 
the distribution of dark matter sub-structure in great detail for the first time.
We will refer to the individual mass clumps resolved here in our discussion of the X-ray analysis below.

\begin{table}
 \caption{Structures identified within our lensing reconstruction.}
 \label{STRUCTURES}
\begin{center}
 \begin{tabular}{cccc}
  \hline
  Name & $x$ & $y$&$M(r<250\textrm{kpc})$ \\
&(\arcsec)&(\arcsec)&($10^{14}~M_{\odot}$)\\
  \hline
Core&$-13^{+11}_{-8}$&$-4^{+6}_{-13}$&$2.24\pm0.55$\\[1ex]
NW&$71^{+11}_{-10}$&$84^{+15}_{-7}$&$1.15\pm0.23$\\[1ex]
W&$177^{+23}_{-17}$&$-30^{+11}_{-15}$&$1.11\pm0.28$\\[1ex]
N&$34^{+23}_{-32}$&$170^{+10}_{-28}$&$0.86\pm0.22$\\
\hline
 \end{tabular}
\end{center}
\medskip
The x-and y-coordinates are provided in arcseconds, relative to the BCG position 
($\alpha_{\textrm{J2000}} =3.58611^{\circ}$, $\delta_{\textrm{J2000}} =-30.40024^{\circ}$). The 
68\% confidence limits
on peak positions are derived from
500 bootstrap realisations for the Core peak, and from the pixel size of the coarse weak lensing mesh in the reconstruction outskirts for
the other mass peaks. Masses assume $h=0.7$ and their 68\% confidence limits are derived from bootstrap realisations as described in the text. 
\end{table}

\section{X-ray analysis}	
\label{XRAY}

The most difficult part of interpreting merging clusters is determining the geometric configuration of the collision, 
such as its impact velocity, impact parameter and angle with respect to the plane of the sky \citep{Markevitch2002}.
For this purpose, X-ray data becomes a crucial addition to lensing measurements.
The location of any shock fronts are revealed in the temperature of the intracluster medium (ICM).
Velocities can be inferred from the density and temperature of intracluster gas (if the merger axis is near the plane of the sky), 
or through direct Doppler measurements (if the merger axis is near the line of sight).

\subsection{Reduction of the Chandra data}

We reanalysed all existing {\sl Chandra} data of Abell~2744 (listed in \citet{Owers2011}), using \texttt{CIAO} 4.3 with the calibration database CALDB 4.4.2.
We cleaned the data using the standard procedure\footnote{http://cxc.harvard.edu/ciao/guides/acis\_data.html} and kept events with grades 0, 2, 3, 4 and 6. 
We removed the {\sl ACIS} particle background as prescribed for `VFAINT' mode, 
and applied gain map correction, together with Pulse Hight Amplitude (PHA) and pixel randomisation. 
Point sources were identified and removed, and the sky background to be subtracted from 
spectral fits was generated from Blank-Sky observations using the {\tt acis\_bkgrnd\_lookup} script. 
We fit spectra using \texttt{XSPEC} V12.6.0q \citep{Arnaud1996} and
we adopt {\sc Vapec} thermal emission model from atomic data in the 
companion Astrophysical Plasma Emission Database \citep{Smith2001}, allowing for the variation 
of several individual elemental abundances during the spectral fittings.
Galactic photoelectric absorption is incorporated using the {\tt wabs} model \citep{Morrison1983}.
Spectral channels are grouped to have at least 20 counts/channel.
Energy ranges are restricted to 0.5--9.5 keV.
Metal abundances are quoted relative to the solar photospheric values of \citet{Anders1989}, 
and the spectral fitting parameter errors are 1-$\sigma$ unless stated otherwise.
 
There is a known reduction of quantum efficiency at energies below 1~keV 
due to a build-up of molecular contaminants on the optical blocking filter
(or on the CCD chips)\footnote{http://cxc.harvard.edu/cal/Links/Acis/\\acis/Cal\_prods/qeDeg/index.html}.
To prevent this from affecting our measurement of low energy line abundances, hydrogen column density and overall gas temperature,
we fix the column density to the cluster's nominal value of 1.6$\times$10$^{20}$~cm$^{-2}$ \citep{Dickey1990}. 
To be conservative, we present results from only those regions of the CCDs best-suited for velocity analysis ({\sl ACIS-I} pointings 8477 and 8557);
for an independent sanity check, we also repeat the analysis with {\sl ACIS-S} pointing 2212.
We exclude pointings 7712 and 7915 because the regions of interest cross multiple CCDs, and interchip gain fluctuations could introduce spurious systematic effects when measuring the redshifts from X-ray spectra. 

Since spectral models are (weakly) degenerate with cluster redshift, it is common for simultaneous 
fitting routines to get stuck in local $\chi^2$ minima before they reach the global best fit. 
We circumvent this via an iterative approach.
We perform initial fits while varying the redshift values within reasonable ranges (via the command STEPPAR in \texttt{XSPEC}). 
We then fix the redshift, and refit full spectral models to infer gas temperature, metal abundances and normalisations. 
Subsequently, we use these best-fit values as inputs in a new fit with the redshift free to vary.
This provides a more reliable estimation of the error on the redshift measurement.
We repeat the process until the best-fit redshift no longer changes between iterations.

\subsection{Previous X-ray observations \& interpretations}

Abell~2744 shows an extremely disturbed X-ray morphology (compare Fig.~\ref{XPROFILE}). 
With 25~ksec of {\sl Chandra} {\sl ACIS-S3} imaging, \citet{Kempner2004} decided it is
the aftermath of a N-S collision between similar mass proto-clusters, at Mach~$>2.6$.
They also found tentative evidence for a cold front in the detached nearby NW `interloper', which they guessed was falling into the main cluster.
\citet{Owers2011} obtained a further $101$~ksec of {\sl ACIS-I} imaging. 
The deeper data revealed the `Southern minor remnant core' (SMRC) to be colder (T$_X\sim$~7.5~keV) than its surroundings, 
with a high temperature region to the SE (T$_X>15$~keV) that they interpreted as a shock front.
They concluded the SMRC had been a low-mass bullet that has passed through the `Northern major remnant core' (NMRC), leaving central tidal debris (CTD).
The pressure ratio of $\sim3$:1 across the shock front corresponds to a sky-projected shock velocity of 2150 km/s (for an average temperature T$_X\sim$~8.6~keV).
\citet{Owers2011} reversed the \citet{Kempner2004} model of the interloper,
concluding that it came originally from the South, has already passed through the main body of 
the host cluster with a large impact parameter, and is now climbing out towards the N-NW.

\citet{Owers2011} also obtained spectra of more than 1200 galaxies with the {\sl Anglo-Australian Telescope} 
multi-fibre {\sl AAOmega} spectrograph, and confirmed the velocity bi-modality of the cluster galaxies.
One component of galaxies near the SMRC has a peculiar velocity of $2300$~km/s; a separate component near the NMRC 
has a peculiar velocity of about $-1600$\~km/s and enhanced metal abundances ($\sim$~0.5 Solar).
Assuming that the Northern and Southern cores have the same velocities as the `apparently' associated galaxy populations, 
they de-projected the velocity of this collision to Mach~3.31 or nearly 5000~km/s. 

An interesting prediction of this scenario is that the intracluster gas should show a strong radial velocity gradient 
of $\sim4000$~km/s or $\Delta~z\sim 0.014$ from North to South.
Since the intracluster medium is enriched with heavy elements, radial velocity measurements could be carried out directly 
by measuring the Doppler shift of emission lines in the X-ray spectrum
\citep{Dupke2001a,Dupke2001,Andersson2004,Dupke2007} or through changes in 
line broadening due to turbulence \citep{Inogamov2003,Sunyaev2003,Pawl2005}. 
The former requires high photon counts within the spectral lines and excellent control of 
instrumental gain but could, in principle, be measured with current X-ray spectrometers.
The latter requires very high spectral resolution that should become available through the 
future {\sl ASTRO-H} and  {\sl IXO} satellites \citep[however, see][]{Sanders2010a,Sanders2011}.

The existing data are not ideal for measurements of ICM velocity structure with high precision due to the variation 
of gain expected from the analysis of multiple observations with different pointings and in different epochs. However, 
the expected radial velocity gradient of 4000km/s, predicted from the optical work of \citet{Owers2011} is higher than the expected
 interchip and intrachip gain variations, so that  it becomes possible to test, even if just for consistency, 
the proposed merger configuration, as we did in this work.
{\sl Chandra} has a good gain temporal stability \citep{Grant2001} and
we shall control for spatial variations by performing resolved spectroscopy 
of multiple cluster regions using {\it the same CCD location} in different {\sl ACIS-I} pointings.

\subsection{Velocity measurements}

\begin{figure}
 \begin{center}
  \includegraphics[width = .48\textwidth]{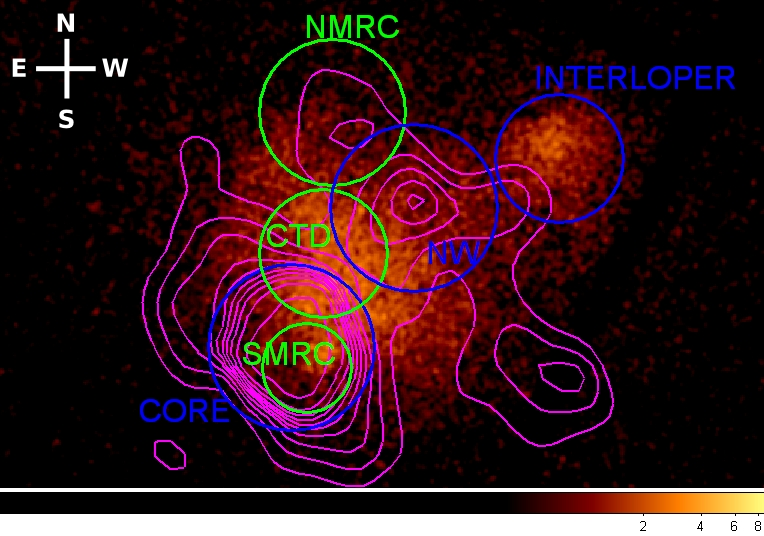}
 \end{center}
\caption{X-ray image of Abell~2744, overlaid with our lensing mass reconstruction (magenta).
The velocity gradient is maximal between three regions of interest named by Owers et al. (2011) 
and circled in green 
(Southern minor remnant core, Northern major remnant core, and Central tidal debris).
A velocity gradient is also detected between the regions of interest from our lensing mass reconstruction, circled in blue
(Core and NW) as well as the interloper. 
}
\label{XPROFILE}
\end{figure}

We perform our velocity analysis twice: for the regions of interest defined by \citet{Owers2011}, then for the substructures prominent in our lensing mass maps.
These are respectively indicated by green or blue circles in Fig.~\ref{XPROFILE}.

We find temperature and metal abundance values for the NMRC, CTD and SMRC regions of
8.29$\pm$0.67 keV and 0.55$\pm$0.18 Solar ($\chi_{\nu}=$1.06 for 234 degrees of freedom);
9.51$\pm$0.47 keV and 0.27$\pm$0.08 Solar ($\chi_{\nu}=$1.03 for 469 degrees of freedom); and 
7.98$\pm$0.76 keV and 1.17$\pm$0.51 Solar ($\chi_{\nu}=$1.01 for 143 degrees of freedom).
These values are derived from simultaneous fits to pointings 8477, 8557 and 2212.
The values are consistent with those obtained through individual spectral fits for each instrument/observation (two independent ACIS-I 
and one ACIS-S). Our temperature and 
abundance measurements are thus consistent with the values found by \citet{Owers2011}. 
Given that the observations were not tailored for velocity measurements 
(\textit{i.e.}\ we cannot exclude temporal or inter-chip gain variations), 
we conservatively include in our velocity error bars secondary $\chi^2$ minima that are separated from the global minimum
at less than 90\% confidence in the velocity measurements.

Our velocity measurements are consistent with the presence of a gradient between the NMRC and SMRC, 
but in the opposite sense to that expected from \citet{Owers2011} interpretation.
Using the ACIS-S, the CCD with the best spectral response, the velocity difference between NMRC and SMRC is found to be $>1500$~km/s at 90\% confidence.

The observations (8477 and 8557) indicate a higher magnitude for the velocity gradient detected. However, 
the central values of the best-fit redshifts in each observation are found to be significantly discrepant, introducing larger uncertainties.
The reason for this discrepancy is not clear, but it is possibly related to the observed focal plane temperature variation of about 0.9C between 
these observations (Catherine Grant Personal Communication).
This explanation is supported by the marginally significant observed systematic difference of $\sim$ 10\%-20\% in the best-fit values of temperatures and abundances measured for the same regions in these different ACIS-I observations. This effect was also noted in \citet{Owers2011} (Owers, M. Personal Communication). 
Therefore, although our analysis is consistent with a high velocity gradient, the direction
of the gradient does not corroborate the merger configuration suggested by \citet{Owers2011}. 

Having the lensing mass reconstruction shown in the previous sections at hand we can study the 
line-of-sight gas velocity mapping for the regions of interest (Core and NW) more precisely.
Using all three exposures, we measure their gas temperatures and metal abundances to be
T$_{X_{Core}}$=10.50$\pm$0.57 keV, A$_{Core}$=0.28$\pm$0.11~Solar with a reduced chi-squared $\chi_{\nu}$=0.90 for 466 degrees of freedom and 
T$_{X_{NW}}$=10.22$\pm$0.54 keV, A$_{NW}$=0.40$\pm$0.09~Solar with $\chi_{\nu}$=0.91 for 477 degrees of freedom.

\begin{figure}
 \begin{center}
   \includegraphics[width = .48\textwidth]{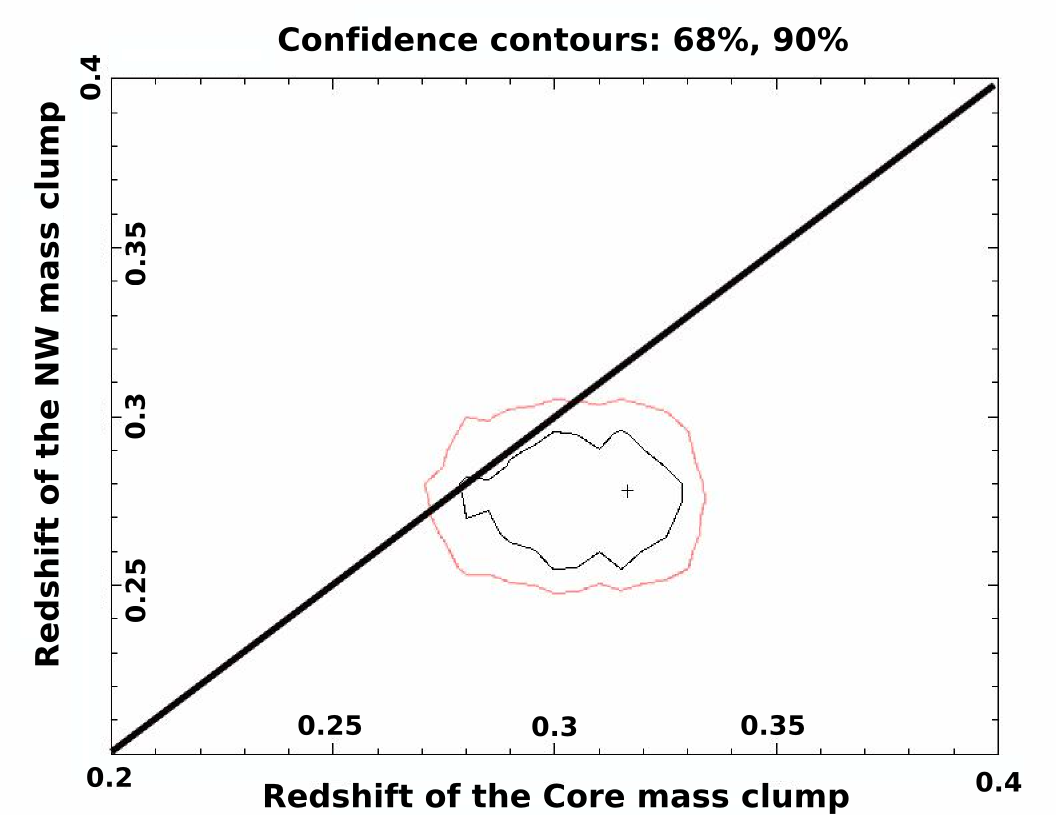}
 \end{center}
\caption{ Simultaneous fits to the redshifts of the Core region and the NW substructure using two {\sl ACIS-I} and one {\sl ACIS-S} pointings.
The two contours correspond to 68\% (inner) and 90\% (outer) confidence levels, and the diagonal line indicates equal redshifts to guide the eye. The Core mass clump appears significantly
redshifted when compared to the NW mass clump. 
}
\label{XREDSHFITS}
\end{figure}

Despite their similar gas properties, these regions show a velocity gradient $>5200$~km/s at near 90\% confidence, 
even including a conservative (1$\sigma$) intra-chip gain variation of 1000~km/s \citep[e.g.][]{Dupke2006} per CCD in the error budget, in quadrature. 
A contour plot of the velocity difference is shown in Fig.~\ref{XREDSHFITS}. 
This result is consistent with the idea that the Southern Core mass is {\it redshifted} with respect to other structures.
Nonetheless, these calculations should be taken with caution due to the uncertainties related to temporal gain variations 
and inter-chip gain variations between {\sl ACIS-I} and {\sl ACIS-S}.
The same analysis using only the {\sl ACIS-I} observations shows the same velocity trend, 
but with a reduction in the absolute velocity difference (an upper limit of $<$4400~km/s at the 90\% confidence level).
Deeper {\sl Chandra} observations, specifically tailored for velocity measurements will be crucial to further reduce uncertainties 
and to better constrain the velocity structure of this cluster.

\section{Interpreting the merger}
\label{INTERPRETATION}
%

Abell~2744 is undoubtedly undergoing a complicated merging process on a large cluster scale. 
Progressively more detailed studies, culminating in our lensing reconstruction (Sec.~\ref{LENSING}) and X-ray analysis (Sec.~\ref{XRAY}),
have only agreed that the merger is more complex than previously thought.
For example, the likely explanation for the gas velocity gradient in the opposite direction to that expected by \citet{Owers2011} is that the NMRC
is \textit{not} the main cluster.
Our lensing mass reconstruction shows that the deepest gravitational potential is by far the Southern `Core' structure, 
which is roughly coincident with Owers et al.'s SMRC but slightly further offset from the Compact Core.
We also find three separate mass concentrations to the North, Northwest and West.
Overlaying the lensing mass reconstruction and X-ray emission in Fig.~\ref{DMXOVERLAY} reveals a complex picture of
separations between the dark matter and baryonic components.
To interpret the sequence of events that led up to this present state, we shall now tour the regions of interest, with more detailed discussions.


\subsection{Core, the massive clump}
According to our lensing analysis, the Core region (lower-left quadrant of Fig.~\ref{DMXOVERLAY}) is by far the most massive structure within the merging system (\textit{c.f.} Tab.~\ref{STRUCTURES}). 
All the strong lensing features can be seen within this clump.
We find no large separation between the distribution of mass and baryonic components. 
The mass peak is centred amongst the bright cluster member galaxies (within $1\sigma$ errors, it is consistent with the position of the BCG)
and only $22\pm12$ arcseconds from a peak of X-ray emission identified by \citet{Owers2011}.
We support the general conclusion of \citet{Owers2011} that the major-merger
in Abell~2744 is similar to that of the \textit{Bullet Cluster} as it would be seen at a large inclination with respect to the plane of the sky.
However, we reverse the ordering of the major and minor mass components.

One can infer constraints on the collisional cross-section of dark matter from the separation between peaks in the lensing and X-ray maps.
For the \textit{Bullet Cluster}, \citet{Markevitch2004} found $\sigma/m < 5~\textrm{cm}^{2}\textrm{g}^{-1}$, and 
for the \textit{Baby Bullet}, \citet{ Bradav2008} found $\sigma/m < 4~\textrm{cm}^{2}\textrm{g}^{-1}$. 
In the Core of Abell~2744, we observe a projected $17\arcsec$ separation that, if the inclination is 
$\sim30^{\circ}$ away from the line-of-sight \citep{Owers2011}, is a physical separation similar to that in the other bullet clusters.
For an order of magnitude analysis, we measure the mean surface-mass density within $150$~kpc of the mass peak 
$\Sigma\simeq0.30^{+0.08}_{-0.07}~\textrm{g}~\textrm{cm}^{-2}$, so that the scattering depth $\tau_{\textrm{s}} = \sigma /m~\Sigma$.
With the assumption $\tau_{s}<1$, which is justified due to the observed dark matter-gas separation, we deduce $\sigma/m < 3\pm1 ~\textrm{cm}^{2}\textrm{g}^{-1}$.
This system may therefore yield one of the tightest constraints on the interaction cross-section of dark matter, based on such analysis.
A full numerical simulation to interpret the cluster configuration would be ideal, 
especially given the uncertainty in the collision angle with respect to the plane of the sky.
Indeed, even tighter constraints ($\sigma/m < 0.7 ~\textrm{cm}^{2}~\textrm{g}^{-1}$) were obtained from the \textit{Bullet Cluster}
by \citep{Randall2008}, who interpreted the offsets between all three cluster components via tailored hydrodynamical simulations.
Additional constraints on the collisional cross-section have also recently been obtained from 
dark matter stripping in Abell~3827 \citep{Williams2011,Carrasco2010} and the ellipticities of dark matter halos \citep{Miralda-Escud'e2002}, with 
implications discussed in \citet{Feng2010}.

\subsection{Northern, the bullet}
Our analysis of the Northern mass substructure (upper-left quadrant of Fig.~\ref{DMXOVERLAY}) confirms the overall North-South merging scenario proposed by several authors in the past.
We find a mass ratio of $\sim2.6$ between the Core and the Northern clump, roughly supporting the 3:1~merging scenario of \citet{Boschin2006},
but we identify the Northern sub-clump as the less massive progenitor.
This conclusion is robust, with no strong lensing features revealed by even our high resolution {\sl HST} imaging in the Northern structure, 
as would have been expected for the reversed mass ordering proposed by \citet{Owers2011}.

X-ray emission in the Northern mass substructure lags behind the dark matter as expected. 
We measure a separation of $\sim30\arcsec$ to the South.
This is a similar separation to that in the core but, due to the lower surface-mass density in this region, constraints on the collisional cross-section are less significant.

\subsection{Northwestern, the ghostly and dark clumps}
By far the most interesting structure is located to the Northwest of the cluster field (upper-right quadrant of Fig.~\ref{DMXOVERLAY}).
Our lensing analysis (Sec.~\ref{LENSING_REC}) shows it to be the second-most massive structure.
A separate region of X-ray emission also lies to the Northwest, called the NW interloper by \citet{Owers2011}.
Furthermore, in 54\% of our weak lensing bootstrap realisation we identified a second peak in the more Western area 
of the NW mass clump (see Fig.~\ref{DMXOVERLAY}), rendering it difficult to say if we indeed see a single, separate dark matter structure
and to derive decisive separation between dark matter, X-ray luminous gas and bright cluster member galaxies.
However, the separation between our different possible NW mass peak positions and the NW interloper is large ($>150$~kpc for NW2).
When compared to the more prominent Eastern mass peak NW1, which is identified in 95\% of the bootstrap realisations, the distance 
is even at least $400$~kpc.

With a de-projected temperature of $\sim5$~keV \citep{Owers2011} the NW interloper should have 
$r_{500} \sim 1.34$~Mpc, \citep{Evrard1996} and $M_{500}\sim4-5\times10^{14} M_{\odot}$ (\textit{e.g.} Fig.~8 of \citet{Khosroshahi2007}). 
Its $0.1-10.0$keV unabsorbed luminosity is $2.5\times10^{44}$~erg/s, consistent with its gas temperature of $4-5$ keV \citep{Khosroshahi2007,D'iaz-Gim'enez2011}.
Assuming a $\beta$ (from a King-like profile) of 0.67, typical for clusters, the clump would have $\sim 0.95\times10^{14} M_{\odot}$ within 250~kpc, 
similar to the N clump, and should have been easily detected in the lensing analysis. 
The interloper thus appears to be an X-ray feature with no associated dark matter or galaxies, and we therefore dub it the `ghost' cluster.

There is also a clear separation between the peak of the NW mass clump and any cluster member galaxies, so we call this the `dark' cluster.
Contours of the lensing mass reconstruction extend towards the West, where indeed we find a pair of giant ellipticals (see Fig.~\ref{DMXOVERLAY}).
However, with the limited resolution of the {\sl VLT} weak lensing reconstruction, it is impossible to tell whether this is a binary mass structure,
though there are clear indications for that in our analysis of this mass clump.

The separation between all three mass components makes this a real puzzle and it should be stressed that such a  peculiar configuration
is observed for the first time. It may pose another serious challenge for cosmological models of structure formation.
One possible explanation was suggested by \citet{Owers2011}, who describe it as a ram-pressure slingshot \citep[e.g.][]{Markevitch2007}.
In this interpretation, after first core-passage, gas initially trails its associated dark matter but, while the dark matter slows down, the gas slingshots past it
due to a combination of low ram-pressure stripping and adiabatic expansion and cooling, which enhances the cold front temperature contrast \citep{Bialek2002}.
There is indeed a clear velocity gradient between the NW interloper and the main cluster core (see Sec.~\ref{XRAY}).
Such effects have also been observed e.g. in Abell~168 \citep{Hallman2004}, in a joint weak-lensing and X-ray study of Abell~754 \citep{Okabe2008}
and in numerical simulations \citep{Ascasibar2006,Mathis2005},
although at a much smaller separations between dark matter and gas.
The more than $100\arcsec$ separation in Abell~2744 suggests either that the slingshot scenario is unlikely or that some amplifying mechanism is in place. 
We shall return to this issue, proposing an interpretation of the entire cluster merger, at the end of this section.

\subsection{Western, the stripped clump}
The Western substructure (lower-right quadrant of Fig.~\ref{DMXOVERLAY}) has not yet been discussed in the literature, but several cluster member galaxies are found in this area.
We find a prominent weak gravitational lensing signal of $\sim1.0\times10^{14} M_{\odot}$ within $250$~kpc.
This should correspond to an X-ray brightness higher than the NW interloper, for the same gas temperature.
However, we detect no X-ray emission.
The best X-ray data ({\sl ACIS-I} pointings 8477 \& 8557) do not cover the region of this lensing signal, but the {\sl ACIS-S} pointing 2212
indicates no excess diffuse gas above the cosmic background and the extended tail of the cluster's outskirts.
To match these observations, the Western clump must have been completely stripped of its ICM, so we dub it the `stripped' clump.

\begin{figure}
 \begin{center}
  \includegraphics[width = .47\textwidth]{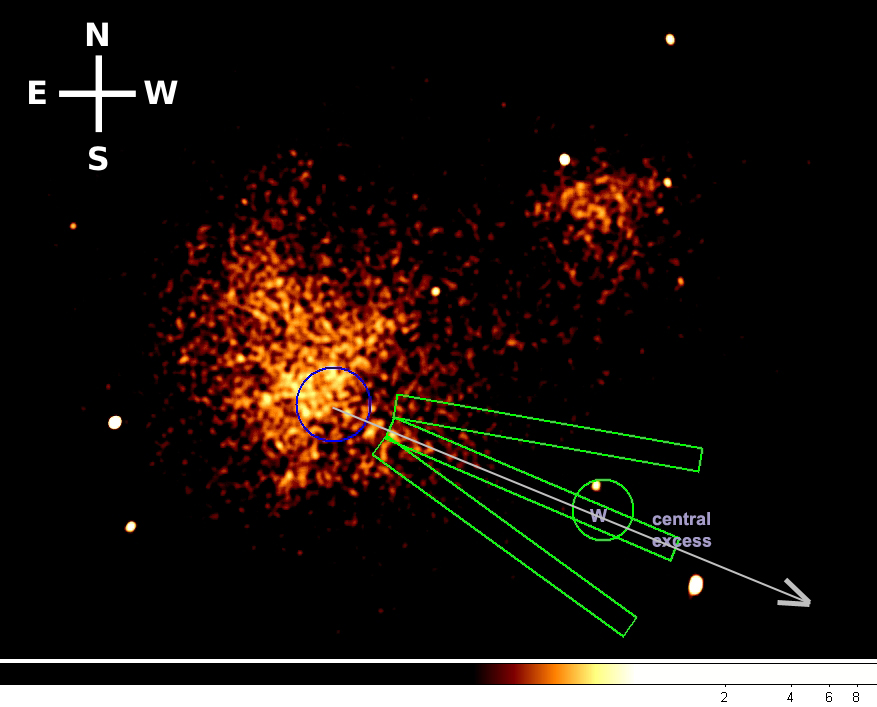}
 \end{center}
\caption{{\sl Chandra} {\sl ACIS-S3} X-ray image of Abell~2744, smoothed with a 3 pixel kernel Gaussian. 
We analyse the cluster's X-ray profile in three rectangular regions, each $150\arcsec$ in length and radiating from the cluster centre (marked by the blue circle). 
We find marginally significant excess X-ray emission in the central rectangular region, which extends towards the Western clump (marked by the green circle).} 
\label{WBOX}
\end{figure}

\begin{figure}
 \begin{center}
  \includegraphics[width = .5\textwidth]{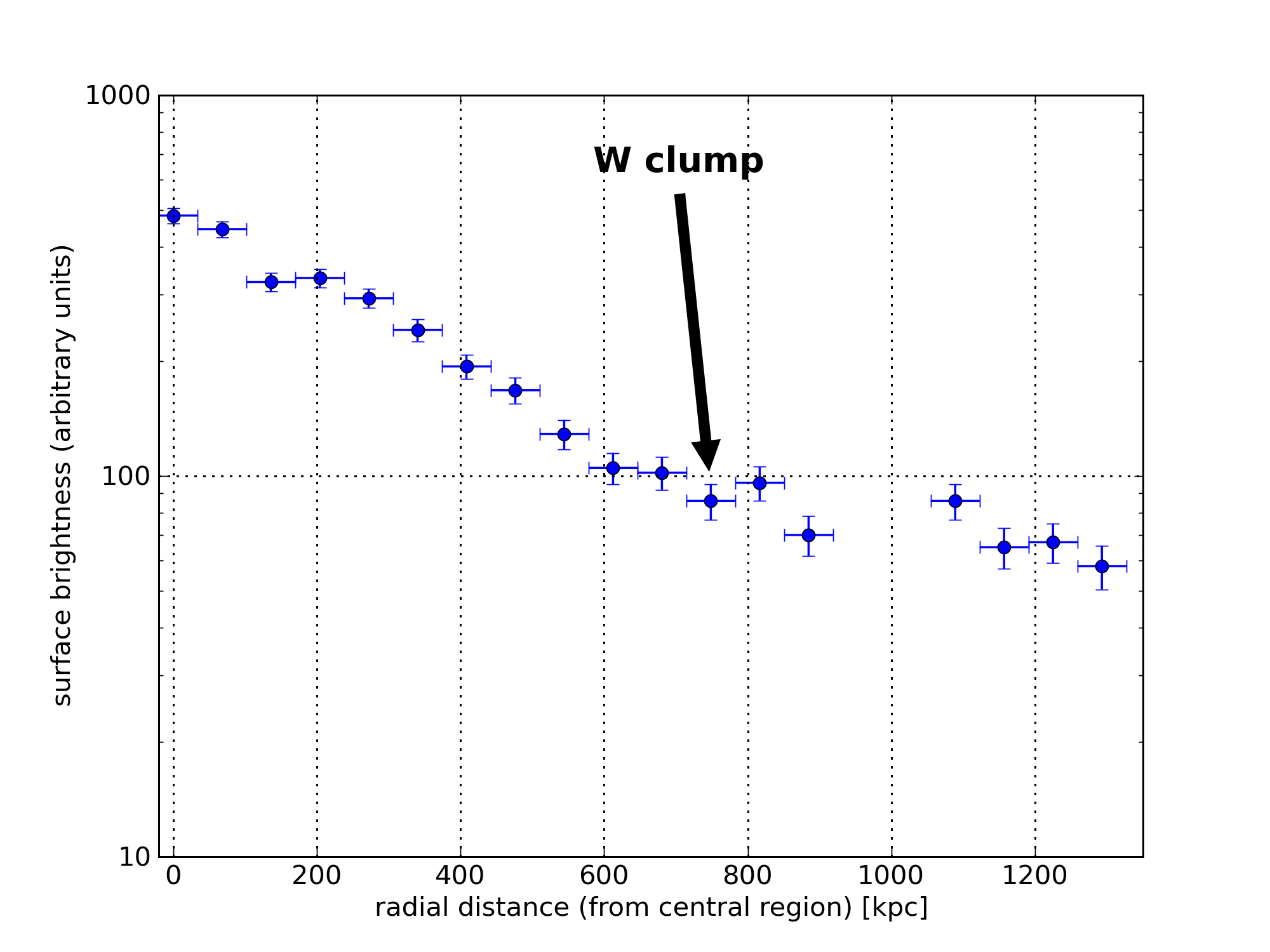}
 \end{center}
\caption{Surface brightness profile (in arbitrary units) of the central rectangular region shown in Fig.~\ref{WBOX}.
The left hand side starts at the X-ray centre, and the arrow denotes the approximate position of the Western mass clump. 
The abrupt change in profile slope at roughly the same location is highlighted by the intersection with the horizontal line.} 
\label{WPROFILE}
\end{figure}

The only slight excess X-ray emission nearby is a faint extension of the main cluster core towards the West \citep[`ridge c' in][]{Owers2011}.
This roughly links the Western clump to the Northern clump, 
and is consistent with remnants stripped by ram-pressure during a secondary merging event (NE--SW), 
almost perpendicular to the main merging event projected in the sky plane. 
To quantify the excess X-ray emission, we measure counts inside three equal-size, non-overlapping regions 
extending from the cluster core to a radial distance of $150\arcsec\approx680$~kpc, as shown in Fig.~\ref{WBOX}.
To prepare the data for this analysis, we divide the image by the exposure map following the standard procedure, remove hot pixels 
and remove point sources using the {\tt dmfilth} routine in \texttt{CIAO}.
We find a marginal excess in the central box (1030$\pm$32 counts), which points to the Western clump,
above the Northern (965$\pm$31 counts) and Southern (876$\pm$29 counts) boxes.
The cluster's surface brightness profile from the core towards the Western clump (along the central box) is shown in Fig.~\ref{WPROFILE}. 
The slope of the profile changes at a radius  $\sim135\arcsec=612$~kpc away from the centre, becoming significantly shallower.
If we take that point of transition as the location of the remaining gas core, we obtain a separation between the gas and dark matter of 30$\arcsec$.
Similar results are found with the shallower observation 7915 using {\sl ACIS-I}. 

We find a slight offset between the peak of the lensing mass reconstruction and the most luminous nearby cluster member galaxies.
However, there is large uncertainty in the position of the lensing peak because this lies outside the {\sl HST} imaging area.
Our weak lensing analysis uses only {\sl VLT} imaging, and there are no strong lensing constraints, so the mass reconstruction has a broad central plateau.
Additional {\sl HST} observations would provide an ideal foundation to better understand the Western area,
which turns out to be playing a significant role in the overall merger.

\subsection{One possible interpretation}
We shall now try to develop a possible explanation of the complex merging scenario that has taken place in Abell~2744.
To recap, we find four mass clumps (Core, N, NW, W) with approximate masses 2.2, 0.8, 1.1, $1.1\times10^{14} M_{\odot}$, respectively. 
The Core, N and W clumps are relatively close to BCGs and hot gas. 
The NW structure, on the other hand, contains separated dark matter, gas and galaxies.

\begin{figure}
 \begin{center}
  \includegraphics[width = .5\textwidth]{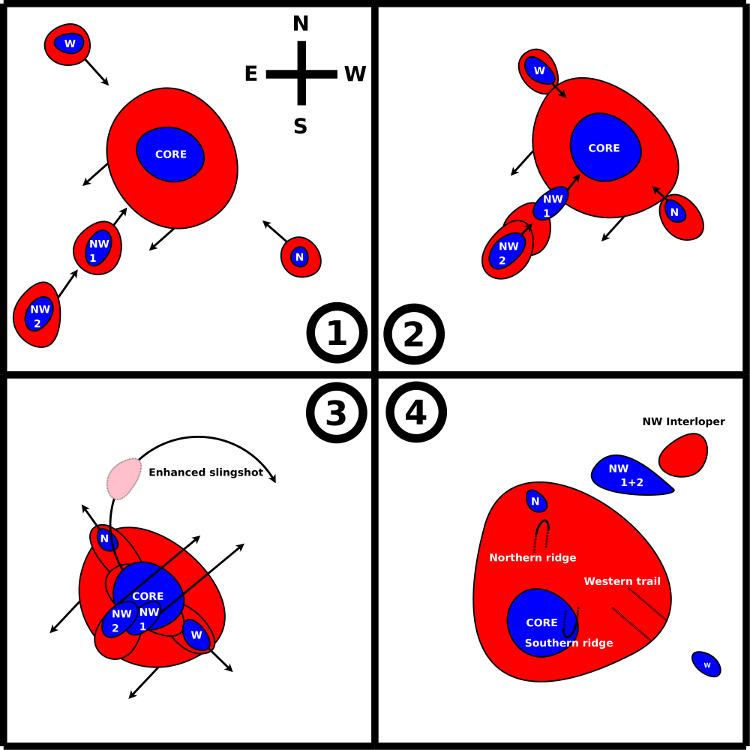}
 \end{center}
\caption{Our proposed merging scenario, illustrated in time-ordered sequence, to explain the current configuration (panel 4).
We suggest that Abell~2744 is the result of a nearly simultaneous double merger: one in the NE-SW direction 
and another in the NW-SE direction, which may even have consisted of three separate structures falling along a filament. Blue colour
shall indicate the innermost dark matter cores of the clumps, where else their respective ICM is shown in red.} 
\label{FOURPANEL}
\end{figure}

We propose that the current configuration is the result of a near simultaneous double merger, as illustrated in Fig.~\ref{FOURPANEL}.
The first merger, in the NE-SW direction, had a characteristic path of $208\arcsec$ (plane of the sky distance between N and W clumps) 
or $\sim0.95$ Mpc (assuming no line of sight velocity component).
The Western clump probably passed closest through the main cluster, as it had its ICM ram-pressure stripped completely.
The second merger, in the SE-NW direction, had a characteristic path of $117\arcsec/\sin(27^{\circ})\approx1.17$ Mpc, 
if we assume the inclination of that merger suggested by \citet{Owers2011} of $27^{\circ}$.
The mergers happened around 0.12--0.15 Gyr ago, with a characteristic velocity of $\sim4000$~km/s 
as indicated by the galaxy velocity difference and the ICM gas velocity measurements.

It is possible that the merger in the SE-NW direction could even have consisted of three initial substructures: 
the Core and two consecutive clumps (with a combined mass (within 250 kpc of each core) of $\sim1.2\times10^{14} M_{\odot}$) falling along a filament.
Those smaller clumps would be accelerated by the gravitational pull of the main cluster 
(plus the Northern and Western clumps, which were merging perpendicularly). 
The ram-pressure slingshot in these clumps could be enhanced by a combination of an initially stronger gravitational field 
and perhaps a posterior reduction in ram-pressure due to the `puff-up' of the gas density due to the recent merger of Core+N+W, 
similar to the density configuration of the main component in the \textit{Bullet Cluster}. 
The combined effect would throw the gas component ahead of its associated dark matter, forming the `ghost' cluster, which is now the interloper. 
The two dark matter clumps left behind would now form the possibly double-centred `dark' clump. 

This unusual scenario fits the current observations. It explains the clear extension of the NW clump towards the West and
provides probably a mechanism to create such a huge separation between dark matter and gas in the NW area. However, the double peak
in the NW clump needs to be confirmed by an additional {\sl HST} pointing to cover its full area. The discovery of a second distinct
mass peak in the NW clump would support our merger scenario and also decrease the large measured separations between dark matter, gas and 
bright cluster galaxies.    
Also the ghostly interloper has not been covered by the current {\sl HST} observations and the Western clump falls between 
chip gaps in the longer {\sl Chandra} exposures, so further observations will still be required.
Our suggestions will also eventually require verification via a set of well-tailored numerical simulations and have to be taken with some
caution at the current stage. Cosmological boxes have to be explored \citep[compare][]{Meneghetti2010,Meneghetti2011} in order to find similar configurations
and detailed, hydrodynamical simulations need to repeat and confirm our findings.


\section{Conclusions}
\label{CONCLUSIONS}

We present a detailed strong lensing, weak lensing and X-ray analysis of the merging cluster of galaxies Abell~2744. 
Earlier studies \citep{Kempner2004,Boschin2006,Braglia2009,Owers2011} concluded that Abell~2744 is undergoing a 
complicated merging event. We find that it is even more complex than previously thought, unleashing a variety of exciting effects.
We dub this merger therefore \textit{Pandora's cluster}.

Deep, three-band {\sl HST}-imaging reveals a variety of strong-lensing features in the core of the cluster.
From our comprehensive strong-lensing modelling of the central mass distribution, 
we identify a total number of not less than 34 multiple images, in 11 multiple-image systems,
together with their respective redshifts. 
The strong lens systems are listed in Tab.~\ref{MSYSTEMS} and 
the finely resolved critical curve of the cluster core is shown in Fig.~\ref{ACS_RGB}.
The Einstein radius of the core is $r_{E}=30\arcsec\pm3\arcsec$. 

We extended the strong lensing information with weak lensing measurements over the whole cluster system ($600\arcsec\times600\arcsec$, or $\sim 2.7$ Mpc on a side). 
The shapes of background galaxies necessary for weak lensing reconstruction were obtained from a comprehensive 
combination of our new {\sl HST/ACS} imaging, the {\sl VLT} data used in \citet{Cypriano2004}, and from additional {\sl Subaru} imaging. 
Our combined strong and weak lensing mass reconstruction (Fig.~\ref{CONV}) resolves a complex structure, with at least four distinct peaks in the
local mass distribution.
The total mass of the cluster is $1.8\pm0.4\times10^{15}M_{\odot}$ within a radius of $1.3$~Mpc, rendering Abell~2744 similar in mass to the \textit{Bullet Cluster}.

{\sl Chandra} X-ray imaging also shows a complex arrangement of substructure.
There are at least four X-ray ridges departing from the X-ray peak, including a Northern (NMRC) and a Southern (SMRC) ridge \citep{Owers2011} . 
Interestingly, none of these coincide with any of the mass clumps found in our lensing mass reconstruction. 
Furthermore, the system also has a separate Northwestern X-ray feature with very low mass, also undetected in our lensing analysis.
Observations of the gas temperature in this Northwestern feature \citep{Kempner2004,Owers2011} show a cold front pointing N-NE
and a shock region ($T_X>15$~keV) in the Southern ridge, for a projected impact velocity of 2150 km/s. 
That projected direction (SE-NW) therefore seems to define the primary merging event. 
However, in contrast to the proposed merging scenario of \citet{Owers2011} , we find the 
Southern ridge to be blueshifted with respect to the Northern ridge. 

We also reveal that the Southern Core is $\sim2.6$ times more massive than the Northern sub-clump --
\citet{Owers2011} had expected this reversed -- 
and that the secondary mass peak is in the Northwest.
The mass ratio thus remains in agreement with general kinematical studies of \citet{Boschin2006}, but swaps the sense of the collision from \citet{Owers2011}.
With this new scenario in hand, we repeated the velocity gradient analysis using the `true' mass clumps (Core and Northwestern). 
These regions exhibit similar gas temperatures and metal abundances, and show evidence for a $>5200$~km/s velocity gradient with the Core region redshifted.
These limits should be taken with caution, since the {\sl Chandra} observations were not tailored specifically for velocity studies
(they were obtained in different CCDs and at different epochs, so could be affected by variations in detector gain),
and there is also some evidence of unknown calibration uncertainty between the two {\sl ACIS-I} pointings taken of the same patch of sky. 
However, our results are consistent with the main merger having a significant component along the line-of-sight, 
with a magnitude and orientation consistent with that seen in the bi-modal distribution of galaxy velocities.
 
We also find evidence for a second merging event, simultaneously with or just before the main merger.
The second merger, along the perpendicular NE-SW axis, was between today's Northern and Western mass peaks.
We think these collided inside the extended halo of the core.
During this dramatic collision, gas in today's Northern `bullet' clump was partially stripped by ram pressure, 
creating a characteristic separation between dark matter and baryons similar to that seen in the \textit{Bullet Cluster}.
The Western `stripped' clump fared worse: all of its gas was removed, strewn into the tidal debris of the Core 
and a faint trail of excess X-ray emission towards its current location where we find just dark matter and galaxies.

The smaller merger may have enabled a curious effect in the main, SE-NW merger.
We postulate that gas in the Core was puffed up by the first collision, reducing ram-pressure stripping during the second.
We also suggest that the main merger could have included two separate subclumps incident along a filament from the SE.
The combined effect would be an enhancement of the `slingshot' effect proposed by \citet{Owers2011}, 
by which the subclumps' gas was accelerated ahead of their dark matter.
This would explain the very large $>35\arcsec$ observed separation between any gas 
(`ghostly' clump or NW interloper), galaxies and dark matter (`dark' clump), as well as the double-peaked morphology of the `dark' clump.
However, this scenario needs further confirmation. 

The interpretation for this spectacular merging system will benefit immensely from additional observations 
and also from numerical simulations that can try to reproduce the new phenomenology shown in this cluster. 
The Western and Northwestern clumps have not been covered by 
{\sl HST} observations and {\sl Chandra} observations of the Western clump are shallow. 
Wide-field {\sl Subaru} imaging in good seeing conditions and in several colours would also be useful to interpret the global environment.
Numerical simulations should be performed to confirm the enhancement of the baryonic slingshot by the 
complicated merger configuration.
The example of the \textit{Bullet Cluster} has shown that the combination of complete lensing and X-ray observations
\citep{Markevitch2004,Bradav2006} with highly resolved hydrodynamical simulations \citep{Springel2007,Randall2008} is a particularly
powerful tool to understand the physics of merging clusters. 
In future work, we shall attempt to repeat this analysis on Abell~2744.
The challenge laid down will be to explain the complicated phenomenology associated with this multiple merger, as well as to
better constrain the collisional cross-section of dark matter, which our rough calculation suggests must be $\sigma/m < 3\pm1~\textrm{cm}^{2}~\textrm{g}^{-1}$,
a tighter constraint than that from a similar analysis of the \textit{Bullet Cluster}, which does not involve numerical simulations.
Therefore, with our own numerical follow-ups we should be able to place similar or even better constraints than e.g. \citet{Randall2008} for
the \textit{Bullet Cluster}.

\section*{Additional affiliations}
$^{11}$Department of Physics and Astronomy, 2130 Fulton St., University of San Francisco, San Francisco, CA 94117, USA\\
$^{12}$Department of Physics and Astronomy, University of British Columbia, 6224 Agricultural Road, Vancouver, BC V6T 1Z1, Canada\\
$^{13}$Instituto de Astrof\'{\i}sica de Andaluc\'{\i}a (CSIC), C/Camino Bajo de Hu\'{e}tor 24, Granada 18008, Spain\\
$^{14}$Department of Theoretical Physics, University of Basque Country UPV/EHU, Leioa, Spain \\
$^{15}$IKERBASQUE, Basque Foundation for Science, 48011 Bilbao, Spain \\
$^{16}$Jet Propulsion Laboratory, California Institute of Technology, 4800 Oak Grove Dr, MS 169-327, Pasadena, CA 91109, USA \\
$^{17}$California Institute of Technology, 1201 East California Blvd, Pasadena, CA 91125, USA\\
$^{18}$Spitzer Science Center, MS 220-6, California Institute of Technology, Pasadena, CA 91125, USA\\

\section*{Acknowledgements}
The authors thank Matt Owers, Catherine Grant and Matthias Bartelmann for useful discussions.
We are particularly grateful to Douglas Clowe, the referee of this work, for his thorough reading of the manuscript. 
His competent comments and suggestions
improved the quality of this work substantially.
JM acknowledges financial support from the Heidelberg Graduate School of Fundamental Physics (HGSFP) 
and by contract research Galaxy clusters of the Baden-W\"{u}rttemberg Stiftung.
All runtime-expensive calculations were performed on
dedicated GPU-machines at the Osservatorio Astronomico di Bologna. 
MM and JM acknowledge financial support from ASI (contracts I/064/08/0, I/009/10/0 and EUCLID-IC) 
and from PRIN INAF 2009.
DC and RD acknowledge partial financial support from grant HST-GO-11689.09-A.
RD also acknowledges support from NASA Grant NNH10CD19C.
RM is supported by STFC Advanced Fellowship \#PP/E006450/1 and ERC grant MIRG-CT-208994.
ESC and LSJ acknowledge support from FAPESP (process ID 2009/07154) and CNPq.
NB and YJT acknowledge support from the Spanish MICINN grant AYA2010-22111-C03-00 and from the Junta de 
Andaluc\'{\i}a Proyecto de Excelencia NBL2003.
This research was carried out in part at the Jet Propulsion Laboratory, 
California Institute of Technology, under a contract with NASA.

\bibliographystyle{mn2e}

\end{document}